\documentclass[prl,twocolumn,showpacs,superscriptaddress]{revtex4}

\usepackage{graphicx}
\usepackage{amsmath}
\usepackage{amssymb}
\usepackage{epsfig}

\renewcommand{\v}[1]{{\bf #1}}

\newcommand{\s}{{\sigma}}

\newcommand{\w}{{\omega}}

\def\eqa{\begin{eqnarray}}
\def\eea{\end{eqnarray}}
\newcommand{\eq}{\begin{equation}}
\newcommand{\ee}{\end{equation}}
\newcommand{\nn}{\nonumber\\}
\newcommand{\Eq}[1]{Eq.~(\ref{#1})}

\renewcommand{\>}{\rangle}

\renewcommand{\Re}{{\rm Re}}
\newcommand{\p}{\partial}

\newcommand{\ua}{\uparrow}
\newcommand{\da}{\downarrow}
\newcommand{\ra}{\rightarrow}

\newcommand{\al}{\alpha}

\newcommand{\del}{\delta}
\newcommand{\Del}{\Delta}
\newcommand{\eps}{\epsilon}

\newcommand{\ga}{\gamma}
\newcommand{\Ga}{\Gamma}

\newcommand{\la}{\lambda}
\newcommand{\La}{\Lambda}

\newcommand{\si}{\sigma}

\usepackage{color}

\begin{document}

\title{Time reversal invariant topological superconductivity in correlated non-centrosymmetric  systems}

\author{Yuan-Yuan Xiang}
\affiliation{National Lab of Solid State Microstructures, Nanjing
University, Nanjing, 210093, China}

\author{Wan-Sheng Wang}
\affiliation{National Lab of Solid State Microstructures, Nanjing
University, Nanjing, 210093, China}

\author{Qiang-Hua Wang}
\affiliation{National Lab of Solid State Microstructures, Nanjing
University, Nanjing, 210093, China}

\author{Dung-Hai Lee}
\affiliation{Department of Physics, University of California at
Berkeley, Berkeley, CA 94720, USA} \affiliation{Materials Sciences
Division, Lawrence Berkeley National Laboratory, Berkeley, CA
94720, USA}


\begin{abstract}
Using functional renormalization group method, we study the
favorable condition for electronic correlation driven time
reversal invariant topological superconductivity in symmetry class
DIII. For non-centrosymmetric systems we argue that the proximity
to ferromagnetic (or small wavevector magnetic) instability can be
used as a guideline for the search of this type of
superconductivity. This is analogous to  the appearance of singlet
unconventional superconductivity in the neighborhood of
antiferromagnetic instability. We show three concrete examples
where ferromagnetic-like fluctuation leads to topological pairing
\end{abstract}


\maketitle

Topological insulators and superconductors have become a focus of
interest in condensed matter physics.\cite{rev1,rev2} These states
are characterized by symmetry protected gapless boundary modes.
The existence of these modes reflects the fact that it is
impossible to deform a topological insulator/superconductor into
its non-topological counterpart without crossing a quantum phase
transition. The free-fermion topological superconductors and
insulators have been classified into ten symmetry
classes\cite{kitaev,ryu}. In each space dimension precisely five
of these classes have representatives. Examples of topological
insulators include the T-breaking integer quantum Hall insulator
(2DEG\cite{2DEG}), and the T-preserving topological insulators in
two and three space dimensions (2D HgTe quantum wells\cite{zhang}
and 3D $Z_2$ topological insulators\cite{Bi2Se3}). Examples of
topological superfluid or superconductor include the T-breaking
$^3$He-A\cite{helium} and Sr$_2$RuO$_4$\cite{Sr2RuO4}(likely), and
T-invariant $^3$He-B\cite{helium}).

In this fast growing field discovering new topological materials
is clearly one of the most important tasks. In this regard
predicting topological superconductors is much harder than
predicting topological insulators. This is  because knowing the
desired Bogoliubov de Gennes (BdG) band structure\cite{schnyder}
only meets half of the challenge. The other half requires the
knowledge the microscopic interactions which favor the desired
quasiparticle band structure as the mean-field theory. There are
many interesting proposals for inducing topologically non-trivial
superconducting pairing via the proximity
effect.\cite{fuliang,sau,alicea} In these proposals, pairing is
artificially induced by a (non-topological) superconductor. The
reason the induced superconducting state is topological is due to
the novel spin-orbit coupled electronic wavefunctions in the
normal state.
A notable exception is the
intriguing proposal that the superconducting state of
Cu$_x$Bi$_2$Se$_3$ is topological.\cite{fu-Berg}

Leaving topology aside, it is extremely challenging to predict
superconductivity itself. This is because the energy scale
involved in Cooper pairing is usually much smaller than the
characteristic energies in the normal state. However in the last
five years various types of renormalization group methods have
been used to compute the {\it effective interaction} responsible
for the Cooper pairing  in iron-based
superconductors.\cite{frg4iron} They lead to a proposal for why
the pairing scale of the pnictides is high: the scattering
channels triggering antiferromagnetism and Cooper pairing have
overlaps. Through these overlaps strong antiferromagnetic
fluctuation enhances superconducting pairing.\cite{wangLee} This
is consistent with the widely known empirical fact: strong
superconducting pairing often occurs when static
antiferromagnetism disappears.

There are two classes (DIII and CI) of time-reversal-invariant
(TRI) topological superconductor in three
dimensions.\cite{kitaev,ryu} They are differentiated by the
transformation properties with respect to time reversal and
particle-hole conjugation. In this paper we will focus on the
so-called class DIII, for it has realization in all space
dimensions. We ask {\it ``under what condition is time-reversal
symmetric topological superconductivity favored?"} We shall argue
that it is near the ferromagnetic (to be precise small wavevector
magnetic) instability. However due to the exponential growth of
computation difficulty with space dimensionality we shall limit
ourselves to two
dimensions. 
We shall also restrict the discussion to systems with a
special type of spin-orbit interaction - the Rashba coupling.
This type of spin-orbit coupling breaks the parity symmetry, hence the superconductors under
consideration are {\it non-centrosymmetric}. For discussions of topological pairing in centrosymmetric systems see, e.g., Ref.\cite{volovik}. Many real
superconducting materials are non-centrosymmetric. Examples include
CePt$_3$Si\cite{ceptsi}, CeRhSi$_3$\cite{cerhsi}, CeIrSi$_3$\cite{ceirsi},  and the superconductivity
found at the interface of LaAlO$_3$ and
SrTiO$_3$\cite{laalo}.

Here are our main results. We present three different mechanisms
for topological pairing. We warm up by studying a one band model
mimicking strongly correlated fermions in continuum. As we shall
discuss this model is relevant to the topological superfluidity in
the B phase of $^3$He. Next we discuss two other different
mechanisms for topological pairing. In each case there is a finite
parameter range where triplet pairing occurs in the presence of
ferromagnetic (or small wavevector magnetic) fluctuation. We
explain why, under such condition, a small Rashba coupling can
induce topological superconductivity. We also explain why
topological pairing does not happen in singlet-dominated
materials. The paper concludes with a guideline for the search of
TRI topological superconductivity in non-centrosymmetric systems.

\begin{figure}
\includegraphics[width=8.5cm]{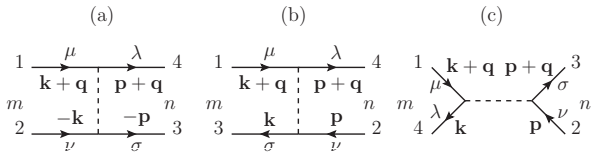}
\caption{A generic 4-point vertex  $\Ga_{1234}$ is rearranged into
$P$-, $C$-, and $D$-channels in (a)-(c), respectively. Here $\v
k,\v q,\v p$ are momenta, $\mu,\nu,\si,\la$ are spin indices, and
$m,n$ denote the form factors. On each side of the diagrams, the
spin (and sublattice) labels are absorbed into the form factor
labels wherever applicable (see the main text).}\label{pcd}
\end{figure}

{\it Method --} Technically this work requires us to generalize
the functional renormalization group (FRG) approach
\cite{frg-general,frg4iron,wangLee,wang} to Hamiltonians without
spin rotation symmetry. In addition, because the necessity to
study small momentum transfer particle-hole scatterings we use a Matsubara frequency rather than momentum cutoff.
All calculations are carried out using  the singular-mode
functional renormalization group (SM-FRG) method.\cite{husemann,wang}

Consider a generic fully-antisymmetized irreducible 4-point vertex
function $\Ga_{1234}$ in $\Psi^\dagger_1 \Psi^\dagger_2
(-\Ga_{1234}) \Psi_3 \Psi_4$. Here $1,2,3,4$ represent momentum
and spin (and sublattice) indices. Figs.\ref{pcd}(a)-(c) are
rearrangements of $\Ga_{1234}$ into the pairing (P), crossing (C)
and  direct (D) channels each characterized by a collective
momentum $\v q$. In each channel the vertex
function is decomposed as \Eq{projection-a} of the Appendix.
There $\{f_m\}$ is a set of orthonormal lattice form
factors.\cite{formfactor} The spin (and sublattice) indices are
contained in the label of the form factors as shown in
Figs.\ref{pcd}(a)-(c). The decomposition  in \Eq{projection-a} is
exact if the form factors are complete, but a few of them are
often enough to capture the leading
instabilities.\cite{husemann,wang} The FRG flow equations for
$P,C$ and $D$ as a function of the cutoff scale $\La$ are given by
Eqs.(\ref{f1}),(\ref{loopint}) and (\ref{f2}) of the Appendix. The
effective interaction in the particle-particle (pp) and
particle-hole (ph) channels are given, respectively, by
$V_{pp}=-P$ and $V_{ph}=C$. [Because of antisymmetry $D$ ($=-C$)
does not yield any new information.] During the FRG flow we
monitor the singular values of the matrix functions $V_{pp/ph}(\v
q)$. The most negative singular values, $S_{pp/ph}$, occur at
special momenta $\v q_{pp/ph}$. While $\v q_{pp}$ is usually zero,
$\v q_{ph}$ can evolve under RG before settling down to fixed
values. The eigen function associated with $S_{pp}$ is used to
construct the gap function. More technical details can be found in
the
Appendix.\\

{\bf A strongly correlated one-band model in continuum limit} --
We consider spin-1/2 fermions hopping on a square lattice. The
Hamiltonian is given by \eqa H&=&\sum_{\v k}\Psi^\dagger_{\v
k}[\epsilon(\v k)\s_0 +\lambda\vec{\gamma}(\v
k)\cdot\vec{\s}]\Psi_{\v
k}\nn&+&U\sum_{i}n_{i\ua}n_{i\da}+V\sum_{\<ij\>}n_in_j=H_0+H_I.\eea
Here $\Psi^\dag=(\psi^\dag_{\ua},\psi^\dag_{\da})$, $\eps(\v
k)=-2t(\cos k_x+\cos k_y)-\mu$ ($t$ is the nearest neighbor
hopping integral and $\mu$ is the chemical potential), $i$ labels
the lattice sites, $n_{i\si}=\psi^\dag_{i\si}\psi_{i\si}$ and
$n_i=\sum_\si n_{i\si}$. In addition, $\s_0$ is the $2\times 2$
identity matrix and $\vec{\si}$ denotes the three Pauli matrices.
In the following we shall set $U=8t$ and $V=-2t$ for the
on-site and nearest neighbor interactions. 
For the Rashba
spin-orbit coupling we consider $\vec{\gamma}(\v k)=(-\sin
k_y,\sin k_x,0)$. 

Combining the time-reversal and point group ($C_{4v}$ in the
present case) symmetries , it can be shown that the Cooper pair
operator $B^\dag=\sum_\v k\Psi^\dag_\v k\Del_\v k\Psi^{\dag
T}_{-\v k}$ takes the form,\cite{noncentrosymmetric} $\Delta(\v
k)=[\phi(\v k)\si_0+\vec{d}(\v k)\cdot\vec{\s}]i\si_2,$ where
$\vec{d}(\v k)$ transforms, under the point group, like the
product of $\phi(\v k)$ and $\vec{\gamma}(\v k)$. In the cases we
have studied, to a good approximation, we can write \eq \Delta(\v
k)=[\phi(\v k)\si_0+\chi(\v k)\hat{\gamma}(\v
k)\cdot\vec{\s}]i\si_2,\label{del2}\ee where $\hat{\ga}(\v
k)=\vec{\ga}(\v k)/|\vec{\ga}(\v k)|$, $\phi(\v k)$ and $\chi(\v
k)$ are even functions of $\v k$ (real up to a global phase) and
transform according to the same irreducible representation of the
point group (for multi-dimensional representations there are
several $\phi$ and $\chi$'s). In Landau theory, $\phi$ and $\chi$
act as order parameters, and can induce each other in the presence
of the Rashba coupling ($\la\neq 0$).

It is important to note that the Rashba term splits each of the
otherwise spin-degenerate Fermi surface into two. The spin split
Fermi surfaces are characterized by  eigen values $\pm 1$ of
$\hat{\gamma}(\v k)\cdot\vec{\s}$. In the case where $\phi(\v k)$
and $\chi(\v k)$ are nodeless, the gap function on the two split
Fermi surfaces will have opposite sign if the magnitudes of
$\chi(\v k)$ dominates over $\phi(\v k)$. It turns out that for
{\it each pair} of Fermi pockets surrounding a TRI $\v k$ point the above sign reversal leads to two
counter-propagating Majorana edge modes . \emph{Thus topological pairing
requires the triplet $\chi$-component to be dominant.} Moreover
sign reversal (in the gap function) on an odd/even pairs of the
spin-split Fermi surfaces (satisfying the condition specified
above) will lead to strong/weak topological superconductivity.

\begin{figure}
\includegraphics[width=8cm]{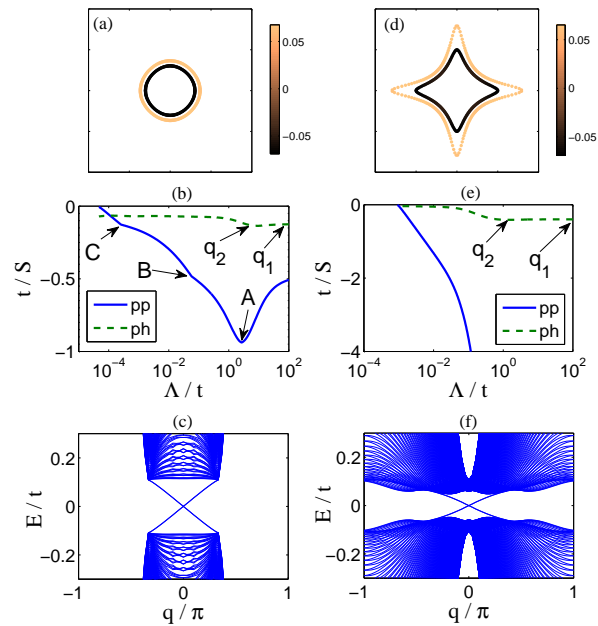}
\caption{(Color online) Left panels: Results for the toy model of
He$^{3}$. (a) The Fermi pockets and the associated gap functions
(gray scale, in units of $t$). The spin-splitting between the
pockets is enlarged for clarity. The box is the zone boundary. (b)
The SM-FRG flow of $S_{pp/ph}$ versus cutoff scale $\La$. Arrows
mark sharp changes in the RG evolution. (c) The low energy BdG
eigen spectrum in a strip (open along $\hat{x}$) as a function of
the momentum $q\hat{y}$. Right panels: the same plots as in the
left panels but for a model whose Fermi surfaces are in proximity
to van Hove singularities.} \label{helium}
\end{figure}

For $\la=0.01t$ and $\mu=-3t$ the spin-split Fermi surfaces are
shown in Fig.\ref{helium}(a). The pockets are small, mimicking the
continuum limit. The form factors used in our SM-FRG extend up to
second neighbors in real space.\cite{formfactor} The RG flow of
$S_{pp/ph}$ are shown in Fig.\ref{helium}(b). During the flow $\v
q_{ph}$ evolves from $\v q_1=(\pi,\pi)$ and settles down at $\v
q_2=0$. By inspecting the spin structure of the $\v q_2$-singular
mode we find it corresponds to ferromagnetic fluctuation. The
leading pairing channel is extended $s$-wave at cutoff energies
above point A (because the bare $V<0$). But at lower cutoff
energies the increased ferromagnetic fluctuation around $\v q_2$
enhances pairing in the triplet channel via their mutual overlaps
(see Appendix). The cusps at A, B and C associated with the
$S_{pp}$ flow is due to the evolution of the leading pairing form
factor. The gap function is determined by the singular mode
associated with $S_{pp}$ at the diverging cutoff scale. The result
is a dominant $\chi$-component together with a much smaller
$\phi$-component. The corresponding gap function on the two Fermi
surfaces is shown in Fig.\ref{helium}(a) (gray scale). A sign
change is clearly visible. According to the established
criterion,\cite{rev2} this pairing state is topological. To verify
this, we calculate the BdG energy spectrum using the obtained
pairing form factor in a strip geometry (open-boundary along
$\hat{x}$). The resulting eigen energies as a function of $q=k_y$
is shown in Fig.\ref{helium}(c). There are two in-gap branches of
Majorana edge modes associated with each edge.

Had we turned off the Rashba coupling, the leading pairing channel
($p$-wave) would be two-fold degenerate (with dominant amplitudes
on 1st neighbor bonds). Under this condition even an infinitesimal
Rashba coupling breaks the degeneracy by linearly recombining the
$p$-waves into $\Delta(\v k)=i\sin k_x\si_0+\sin k_y\si_3$, or
$\chi(\v k)=|\vec{\ga}(\v k)|$ in Eq.(\ref{del2}), leading to a
gap function $\pm \chi(\v k)$ on the infinitesimally split Fermi
surfaces. This gap function has the same symmetry as the two
dimensional version of the $^3$He B phase. In fact there are
strong similarities between the model (and its properties)
described above and the B phase of $^3$He. For example  The small
filling fraction enables this model to describe the continuum
limit. The strong on-site repulsion mimics the short-range strong
repulsive correlation in $^3$He, while the weaker nearest neighbor
attraction is a caricature of the tail of the Lennard-Jones
potential. The enhanced ferromagnetic fluctuation and the
resulting triplet pairing is consistent with the pairing mechanism
described by Anderson and Brinkman\cite{ab}. In addition, the
Rashba coupling plays
a similar role as the parity -invariant spin-orbit interaction in $^3$He: they both lift the degeneracy in the pairing channel.\\

{\bf Topological pairing in the vicinity of van Hove
singularity--} In this section we demonstrate that topological
pairing can be enhanced when the Fermi surfaces are close to van
Hove singularities.  Here we set $V=0$ (so that the interaction is
{\it purely repulsive}), and add a 2nd neighbor hopping $t'$ so
that $\eps(\v k)=-2t(\cos k_x+\cos k_y)-4t'\cos k_x\cos k_y-\mu$.
For $t'=-0.475t$, $\mu=-2t$ and $\la=0.01t$, the spin-split Fermi
surfaces are shown in Fig.\ref{helium}(d). They are pointy along
$\hat{x}$ and $\hat{y}$ reflecting the existence of saddle points
(van Hove singularities) on the Brillouin zone boundary. These
features lead to enhanced ferromagnetic correlations via the
Stoner mechanism. As a consequence triplet pairing is enhanced and
the diverging scale of $S_{pp}$ is raised. These are shown
in Fig.\ref{helium}(e) for $U=2.5t$. The arrows associated with
the $S_{ph}$ flow record the $\v q_{ph}$ evolution from $\v
q_1=(\pi,\pi)$ to $\v q_2=0$. As in the previous section, this
implies ferromagnetic correlations at low energies. The final gap
function on the Fermi surfaces are shown in Fig.\ref{helium}(d)
(gray scale), again with desired sign reversal. Such a pairing
function leads to the BdG energy spectrum shown in
Fig.\ref{helium}(f) in a strip geometry. The edge modes are
apparent.\\

{\bf Topological pairing enhanced by inter-pocket scattering --}
In this section we show a third route to topological pairing.
In this case pairing is triggered by inter-Fermi surface
scattering in a way similar to the pairing in the
pnictides\cite{frg4iron,wangLee}.

\begin{figure}
\includegraphics[width=8.5cm]{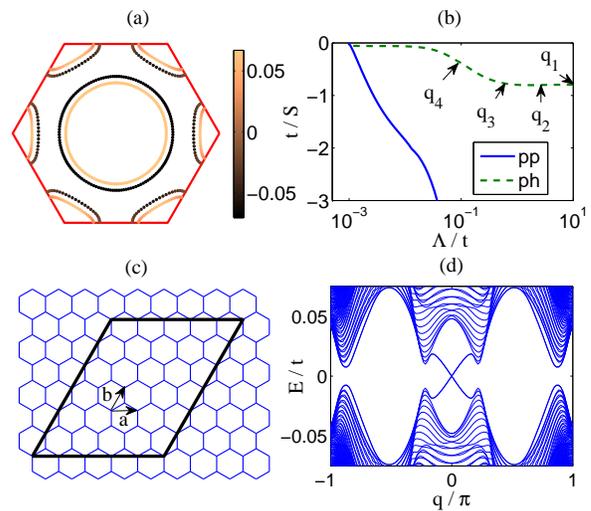}
\caption{(Color online) (a) The Fermi pockets and the associated
gap functions (gray scale, in units of $t$). The spin-splitting
between each pair of Fermi pockets is enlarged for clarity. The
hexagon is the zone boundary. (b) The SM-FRG flow of $S_{pp/ph}$
versus the cutoff scale $\La$. Arrows mark sharp changes of $\v
q_{ph}$ during the RG flow. (c) A strip (marked by the thick
lines) open along $\v a$ and periodic along $\v b$ directions ($\v
a$ and $\v b$ are primitive lattice vectors). (d) The low energy
BdG eigen spectrum for (c) as a function of the conserved momentum
$q$ along $\v b$.}\label{honeycomb}
\end{figure}

Consider a honeycomb lattice. The single particle Hamiltonian is
given by \eqa H_0=&&-\sum_{i \del}\Psi_i^\dag
t_\del\Psi_{i+\del}-i\la\sum_{i\del_{nn}}\Psi^\dag_i
(\hat{z}\times\vec{\del}_{nn}\cdot\vec{\s})\Psi_{i+\del_{nn}}\nn &&
-\mu\sum_i\Psi^\dag_i\Psi_i.\eea Here $i$ labels lattice sites,
$\del$ runs over the 1st and 2nd neighbor bonds, with
$t_\delta=t,t'$. The spin-dependent hopping, the Rashba term, is
limited to the nearest neighbor bonds $\del_{nn}$. 
Choosing a lattice site as the origin, the point group is
$C_{3v}$. For the SM-FRG calculation, we choose the form factors
up to the 2nd neighbors.\cite{formfactor} (Since the honeycomb
lattice has two sites per unit cell the labels of the form factors
in Fig.\ref{pcd} include the sublattice indices.\cite{wang}.) The
Fermi surfaces for $t'=0.357t$, $\la=0.02t$ and $\mu=1.664t$ are
shown in Fig.\ref{honeycomb}(a). There are a few interesting
features of the band structure that are worth noting (1) The Fermi
surfaces encircle either the zone center ($\Gamma$) or the zone
corners (K and K$'$). However only $\Gamma$ is TRI, hence according to Ref.\cite{rev2} only the
$\Gamma$-Fermi surfaces are topologically relevant. (2) The
$\Gamma$ and K-pockets have close by segments, hence allow small
momentum transfer particle-hole scattering. If such scattering is magnetic, it corresponds to
nearly ferromagnetic fluctuations, hence can induce triplet and
topological pairing.

In the following we show for $U=1.26t$ this is exactly what
happens. During the RG flow shown in Fig.\ref{honeycomb}(b), the
strength of $S_{ph}$ increases and $\v q_{ph}$ evolves from $\v
q_1=(0.667,1.152)\pi$ to $\v q_2=0$, $\v q_3=(0.250, 0.048)\pi$
and finally settles down at $\v q_4=(0.333, 0.192)\pi$. We have
checked that $\v q_4$ corresponds to the scattering between near
by parallel segments between the $\Gamma$ and K pockets.
Inspection of the spin structure of the singular mode associated
with $\v q_{2,3,4}$ reveals that they corresponds to spin
fluctuations. As such fluctuations are enhanced, they causes
$S_{pp}$ to grow in magnitude and eventually diverge at a
relatively high critical scale. The resulting gap function is
shown in Fig.\ref{honeycomb}(a) in gray scale. It is {\it fully
gapped on all Fermi surfaces}, and have opposite sign on each pair
of spin-split pockets. Since the K-pockets are topologically
irrelevant, the sign change between the $\Gamma$ Fermi surfaces
implies the pairing is strong-topological. To verify this we
consider a strip schematically shown in Fig.\ref{honeycomb}(c). It
is open along $\v a$ and periodic along $\v b$ directions. The BdG
energy spectrum as a function of the momentum $q=\v k\cdot
\hat{b}$ is shown in Fig.\ref{honeycomb}(d). There are two
branches of Majorana modes at each edge.

Thus in all of the above examples we have seen small momentum
magnetic fluctuations $\Rightarrow$ degenerate triplet pairing,
and degenerate triplet pairing + Rashba coupling $\Rightarrow$
topological pairing. The fact that ferromagnetic fluctuations
enhance triplet pairing has a long history. These include the
works on the pairing of $^3$He,\cite{helium}, and the extension of
the Kohn-Luttinger theorem to $p$-wave pairing for 2D and 3D
electron gas in the dilute limit.\cite{fl,chub} For lattice
systems, a 2D Hubbard model with large enough 2nd neighbor hopping
has been shown to exhibit $p$-wave pairing for small band
fillings.\cite{2dh}

Finally, if pairing is singlet
in the absence of spin-orbit interaction a weak Rashba coupling
only induces a small triplet component, hence is insufficient to
induce the desired sign change in the gap function. Of course this
does not rule out the possibility of topological pairing in the
presence of strong spin-orbit interaction.

In conclusion through the study of the above three, and many other
not included, examples we conclude that TRI topological
superconductivity in symmetry class DIII should occur in systems
close to the ferromagnetic (or small wavevector magnetic)
instability. Bandstructure wise, in the absence of Rashba
coupling, these systems should have an odd number of
spin-degenerate Fermi pockets (each enclosing a TRI momentum) in order for strong topological pairing to occur.\\

\acknowledgments{QHW acknowledges the support by NSFC (under grant
No.10974086, No.10734120 and No.11023002) and the Ministry of
Science and Technology of China (under grant No.2011CBA00108 and
2011CB922101). DHL acknowledges the support by the DOE grant
number DE-AC02-05CH11231.}

\section{Appendix}

In the supplementary material, we provide the technical details of
the SM-FRG method.\cite{wang}.

\begin{figure}
\includegraphics[width=8.5cm]{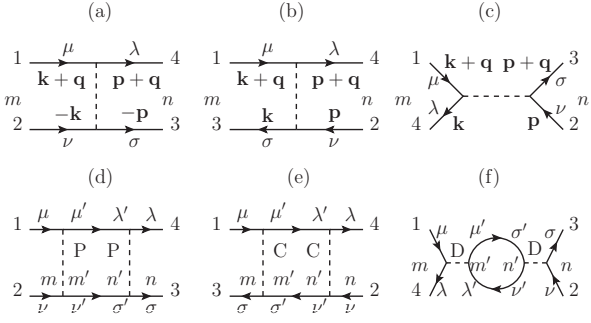}
\caption{A generic 4-point vertex $\Ga_{1234}$ is rearranged into
$P$-, $C$-, and $D$-channels in (a)-(c), respectively. Here $\v
k,\v q,\v p$ are momenta, $\mu,\nu,\si,\la$ denote spins, and
$m,n$ denote the form factors. On each side of the diagrams, the
spin (and sublattice) labels are absorbed into the form factor
labels wherever applicable. The one-loop diagrams that contribute
to $\p P$, $\p C$ and $\p D$ are shown in (d)-(f), respectively.}
\label{diagrams}
\end{figure}

We begin by reviewing the definition of the vertex functions used
in the main text. Consider a generic fully-antisymmetized
irreducible 4-point vertex function $\Ga_{1234}$ in
$\Psi^\dagger_1 \Psi^\dagger_2 (-\Ga_{1234}) \Psi_3 \Psi_4$. Here
$1,2,3,4$ represent momentum and spin (and sublattice) indices.
Figs.\ref{diagrams}(a)-(c) are rearrangements of $\Ga_{1234}$ into
the pairing (P), crossing (C) and  direct (D) channels each
characterized by a collective momentum $\v q$. The rest momentum
dependence of the vertex function can be decomposed as, \eqa &&
\Ga_{\v k+\v q,-\v k,-\v p,\v p+\v q}^{\mu\nu\si\la}\ra
\sum_{mn}f_m^*(\v k)P_{mn}(\v q)f_n(\v p),\nn && \Ga_{\v k+\v q,\v
p,\v k,\v p+\v q}^{\mu\nu\si\la}\ra \sum_{mn}f_m^*(\v k)C_{mn}(\v
q)f_n(\v p),\nn&& \Ga_{\v k+\v q,\v p,\v p+\v q,\v
k}^{\mu\nu\si\la}\ra \sum_{mn}f_m^*(\v k)D_{mn}(\v q)f_n(\v
p).\label{projection-a} \eea Here $\{f_m\}$ is a set of
orthonormal lattice form factors. The spin (and sublattice)
indices are contained in the label of the form factors as shown in
Figs.\ref{diagrams}(a)-(c). The decomposition in \Eq{projection-a}
is exact if the form factors are complete, but in practice a few
of them are often enough to capture the leading
instabilities.\cite{husemann,wang} Because of full antisymmetry,
the matrices $C$ and $D$ satisfy $D=-C$, and are therefore not
independent. In the following $D$ is used for bookkeeping purpose.

Ignoring the spin and sublattice labels for the moment, the form
factors are given by  \eq f_m(\v k)=\sum_{\v r} f_m(\v r)
\exp(-i\v k\cdot\v r),\ee where $f_m(\v r)$ transforms according
to an irreducible representation of the point group, and $\v r$ is
the bond vectors connecting the two $\Psi$'s (or two
$\Psi^\dag$'s) in Fig.\ref{diagrams}(a) and one $\Psi$ and one
$\Psi^\dag$ in Fig.\ref{diagrams}(b) and (c). In our calculation
we choose form factors up to the 2nd neighbor bonds. We have
checked that longer range form factors does not change the results
qualitatively. To be specific, for square lattice, the real-space
form factors we used are 1) $f_1=1$ for on-site; 2) $f_2=1/2$,
$f_3=(1/2)\cos 2\theta_\v r$, $f_4=\sqrt{1/2}\cos\theta_\v r$, and
$f_5=\sqrt{1/2}\sin \theta_\v r$ for 1st neighbors, where
$\theta_\v r$ is the azimuthal angle of $\v r$; 3) $f_6=1/2$,
$f_7=(1/2)\sin 2\theta_\v r$, $f_8=\sqrt{1/2}\cos(\theta_\v
r-\pi/4)$ and $f_9=\sqrt{1/2}\sin (\theta_\v r-\pi/4)$ for 2nd
neighbors.  For hexagonal lattices, the form factors we used are
1) $f_1=1$ for on-site; 2) $f_2=\sqrt{1/3}$, $f_3=\sqrt{2/3}\cos
\theta_\v r$ and $f_4=\sqrt{2/3}\sin \theta_\v r$ for 1st
neighbors; 3) $f_5=\sqrt{1/6}$, $f_6=\sqrt{1/3}\cos\theta_\v r$,
$f_7=\sqrt{1/3}\sin\theta_\v r$, $f_8=\sqrt{1/3}\cos 2\theta_\v
r$, $f_9=\sqrt{1/3}\sin 2\theta_\v r$, $f_{10}=\sqrt{1/6}\cos
3\theta_\v r$ for 2nd neighbors. Notice that the 1st neighbor
bonds stem from different sublattices are negative to each other.

In the case where sublattices are involved, the form factor label
$m$ also includes the sublattice indices associated with the two
$\Psi$'s (or $\Psi^\dag$'s),  or the  $\Psi$ and $\Psi^\dagger$.
However, once $\v r$ is fixed only one of these sublattice indices
is independent. We include the independent sublattice index in the
form factor label, $(m,a)\ra m$. Here $a$ labels, e.g., the
fermion field 1 or 4 in Fig.\ref{diagrams}(a), 1 or 4 in (b), and
1 or 3 in (c). The sublattice index is an independent label
because point group operations do not mix sublattices when the
origin is chosen to be a lattice site.


The total number of form factors $N$ in a calculation is
determined by the number of real space neighbors, the number of
sublattices and the four spin combinations
$(\mu,\nu)=\ua\ua,\ua\da,\da\ua,\da\da$ associated with two $\Psi$
(P channel) or the $\Psi$ and $\Psi^\dagger$ (C and D channels).
Thus $P$, $C$ and $D$ are all $N\times N$ matrix functions of
momentum $\v q$.

The Feynman diagrams associated with one-loop contributions to the
flow of the irreducible 4-point vertex function are given in
Fig.\ref{diagrams}(d)-(f). They represent the partial changes $\p
P$, $\p C$ and $\p D$, respectively. (Notice that the three
diagrams in Fig.\ref{diagrams}(d)-(f) become the usual five
diagrams in the spin-conserved case.) The internal Greens
functions are convoluted with the form factors 
hence in matrix form, \eqa &&\p P/\p \La = P \chi'_{pp} P/2,\nn &&
\p C/\p\La = C \chi'_{ph} C,\nn && \p D/\p\La = -D \chi_{ph}'
D,\label{f1}\eea where we have suppressed the dependence of the
collective momentum $\v q$, and
\begin{widetext} \eqa
(\chi'_{pp})_{mn}&&=\frac{\p}{\p\La}\int\frac{d\w_n}{2\pi}\int\frac{d^2\v
p}{S_{BZ}}f_m(\v p)G(\v p+\v q,i\w_n)G(-\v p,-i\w_n)f_n^*(\v
p)\theta(|\w_n|-\La) \nn &&=-\frac{1}{2\pi}\int\frac{d^2\v
p}{S_{BZ}}f_m(\v p)G(\v p+\v q,i\La)G(-\v p,-i\La)f_n^*(\v p)\ \
+(\La\ra -\La),\nn
(\chi'_{ph})_{mn}&&=\frac{\p}{\p\La}\int\frac{d\w_n}{2\pi}\int\frac{d^2\v
p}{S_{BZ}}f_m(\v p)G(\v p+\v q,i\w_n)G(\v p,i\w_n)f_n^*(\v
p)\theta(|\w_n|-\La) \nn &&=-\frac{1}{2\pi}\int\frac{d^2\v
p}{S_{BZ}}f_m(\v p)G(\v p+\v q,i\La)G(\v p,i\La)f_n^*(\v p)\ \
+(\La\ra -\La),\label{loopint} \eea \end{widetext} where $G$ is
the free fermion Greens function, $S_{BZ}$ is the total area of the
Brillouine zone. Here $\La>0$ is the infrared cutoff of the
Matsubara frequency $\w_n$. As in usual FRG implementation, the
self energy correction and frequency dependence of the vertex
function are ignored.

Since $\p P$, $\p C$ and $\p D$ come from independent one-loop
diagrams, they contribute independently to the full $d\Ga_{1234}$,
which needs to be projected onto the three channels. Therefore the
full flow equations are given by, formally, \eqa && dP/d\La = \p
P/\p\La + \hat{P} (\p C/\p\La+\p D/\p\La),\nn && dC/d\La =\p
C/\p\La + \hat{C}(\p P/\p\La + \p D/\p\La), \nn && dD/d\La = \p
D/\p\La +\hat{D}(\p P/\p\La + \p C/\p\La), \label{f2} \eea where
$\hat{P},\hat{C}$ and $\hat{D}$ are the projection operators in
the sense of Eq.~(\ref{projection-a}). Here we have used the fact
that $\hat{K}(\p K)=\p K$ for $K=P,C,D$. In Eq.~(\ref{f2}) the
terms preceded by the projection operators represent the overlaps
of different channels. For two channels to overlap, the spatial
coordinates of all four fermion fields must lie within the range
set of the form factors. In the actual calculation the projections
in Eq.(\ref{f2}) are preformed in real space.

The effective interaction in the particle-particle (pp) and
particle-hole (ph) channels are given, respectively, by
$V_{pp}=-P$ and $V_{ph}=C$. By singular value decomposition, we
determine the leading instability in each channel,\eqa
V_{X}^{mn}(\v q_X)=\sum_\al S_X^\al\phi_X^\al(m)\psi_X^\al(n),\eea
where $X=pp,ph$, $S_X^\al$ is the singular value of the $\al$-th
singular mode, $\phi_X^\al$ and $\psi_X^\al$ are the right and
left eigen vectors of $V_X$, respectively. We fix the phase of the
eigen vectors by requiring $\Re
[\sum_m\phi_X^\al(m)\psi_X^\al(m)]>0$ so that $S_X^\al<0$
corresponds to an attractive mode in the X-channel.

In the pp-channel $\v q_{pp}=0$ corresponds to the zero
center-of-mass momentum Cooper instability. The matrix gap
function $\Delta_{\v k}$ in the spin and sublattice basis is
determined as follows. A singular mode $\phi_{pp}^\al$ leads to a
pair operator (in the momentum space), \eq \Psi^\dag_\v k\Del_\v
k\Psi^{\dag T}_{-\v k}=\sum_{m=(m,a,\mu,\nu)} \psi_{a\mu}^\dag (\v
k)\phi_{pp}^\al(m)f_m(\v k)^*\psi_{a_m\nu}^\dag (-\v k).\ee Here
$a$ is the independent sublattice index, and $a_m$ is the second
sublattice index determined by $a$ and $m$ as discussed earlier,
and $\mu,\nu$ are spin indices. The parity of $\Del_\v k$ under
space inversion determines the singlet and triplet components. The
gap function in the band eigen basis can be determined by the
unitary transformation \eq \tilde{\Psi}^\dag_\v k=\Psi^\dag_\v k
U^\dag_\v k,\label{ut}\ee where the columns of $U^\dag_\v k$ are
the Bloch states $\{|\v k,n\>\}$ ($n$ is the band index). Under
\Eq{ut} the pairing matrix transforms into \eq \tilde{\Del}_\v k=
U_\v k\Del_\v k U_{-\v k}^T.\label{gt}\ee In the weak coupling
case (i.e., when the magnitude of the superconducting gap is much
smaller than the bandwidth), only the diagonal part of
$\tilde{\Del}$ (i.e., intra-Fermi surface pairing) is important.
Since \Eq{gt} involves Bloch states at two different momenta, the
phases of the associated  Bloch states enters $\tilde{\Del}$.
Since there is time-reversal symmetry we fix the Bloch state phase
at $\v k$ and $-\v k$ by demanding $\hat{T}|\v k,n\>=|-\v k,n\>$
and $\hat{T}^2|\v k,n\>=-|\v k,n\>$, where $\hat{T}=i\si_2 K$ is
the time-reversal operator.

In the particle-hole channel, we calculate the singular values
associated with $V_{ph}(\v q)$ at all momenta $\v q$. Unlike the
Col[oooper channel, the most negative singular value can occur at
non-zero momentum $\v q_{ph}$.  The associated particle-hole
operator is given by \eqa &&\Psi^\dag_{\v k+\v q} \Pi_\v k\Psi_\v
k=\nn&&\sum_{m=(m,a,\mu,\nu)} \psi_{a\mu}^\dag (\v k+\v q
)\phi_{ph}^\al(m)f^*_m(\v k)\psi_{a_m\nu} (\v k).~\eea  Usually
the on-site form factor dominates in the particle-hole channel. By
inspecting the spin structure of the on-site form factor one can
easily determine whether the instability is charge or spin like.


\begin{references}
\bibitem{formfactor}
In real space (ignoring other labels), the form factors \{ $f_m(\v
r)\}$ are irreducible representations of the point group, where
$\v r$ is the bond vector connecting lattice sites. Bond vectors
connected by point group operations form a support for the form
factor. For a given support, the number of independent form
factors equals the number of independent bonds therein.
\end{references}
\end{document}